\documentclass[11pt,showpacs,preprintnumbers,amsmath,amssymb]{revtex4}

\usepackage{graphicx}
\usepackage{dcolumn}
\usepackage{bm}
\usepackage{mathrsfs}
\usepackage{amssymb}
\usepackage{amsmath}
\usepackage{amsfonts}
\usepackage{amsfonts,enumerate}
\usepackage{pifont}
\usepackage{enumerate}
\usepackage{graphics}
\usepackage{verbatim}
\usepackage{amsthm}

\begin{document}


\title{Necessary and sufficient condition for a realistic theory of quantum systems}

\author{Zeqian Chen}
\email{zqchen@wipm.ac.cn}
\affiliation{%
Wuhan Institute of Physics and Mathematics, Chinese Academy of Sciences, 30 West District, Xiao-Hong-Shan, Wuhan 430071, China}%


\date{\today}

\begin{abstract}
We study the possibility to describe pure quantum states and evens with classical probability distributions and conditional probabilities and show that the distributions and/or conditional probabilities have to assume negative values, except for a simple model whose realistic space dimension is not smaller than the Hilbert space dimension of the quantum system. This gives a negative answer to a question proposed by Montina [Phys.Rev.Lett.{\bf 97}, 180401 (2006)] whether or not does there exist a classical theory whose phase-space dimension is much smaller than the Hilbert space dimension for any quantum system. Thus, any realistic theory of quantum mechanics with nonnegative probability distributions and conditional probabilities requires a number of variables grows exponentially with the physical size.
\end{abstract}

\pacs{03.65.Ta, 02.50.Cw}
\maketitle

A peculiar aspect of the standard quantum formalism is linear superposition of quantum states. This characteristic is at the basis of some famous paradoxes, such as the Einstein-Podolsky-Rosen (EPR) paradox \cite{EPR}, Schr\"{o}dinger's cat \cite{S}, and Feymann's which-way experiment \cite{Feymann}. Since the birth of quantum mechanics,
physicists began to investigate the possibility of its description in terms of classical probabilities, such as the Wigner function \cite{Wigner}, before Kolmogorov \cite{Kol} presented the axiomatic system of classical probability theory. The Wigner function has some properties of a classical probability distribution in phase space, but in general it can take negative values. In order to overcome the sign problem of the Wigner function, some alternative non-negative probability distributions were introduced, such as the $Q$ function and the $P$ function \cite{HS}. However, also in these cases, a classical interpretation is not possible in general.

Since quantum mechanics has a statistical interpretation, there is no {\it a priori} reason to exclude the possibility to describe the quantum systems as realistic statistical ensembles. Indeed, a completely equivalent alternative to the Copenhagen interpretation was established by Bohm \cite{Bohm}. Although Bohm mechanics provides a realistic description of quantum systems, it uses the same mathematical tools of the standard approach as the wave function, so it differs only in the interpretation and does not solve, for example, the problem of exponential complexity as pointed out in \cite{M}. More recently, a classical, i.e deterministic theory underlying quantum mechanics was presented by 't Hooft \cite{Hooft}. This brings forward an important new approach towards the interpretation of quantum mechanics, but there are many questions that have not yet been answered, such as how to construct explicit models in which energy can be seen as extrinsic and how to gain more understanding of the phenomenon of interference.

In this article, from a generic probability-theoretic point of view we analyze the possibility to have a realistic description for experiments involving general states and measurements. In Kolmogorov's axiomatic system of classical probability theory there are three basic elements: the nonnegativity of probability, the dispersion-free property of conditional probabilities, and the addition law of probability. As is well known, discarding the nonnegativity of probability Wigner \cite{Wigner} has obtained a realistic phase-space description of quantum mechanics via the Wigner distribution. Also, a generic realistic theory for quantum systems without the dispersion-free property of conditional probabilities has been presented in \cite{BB}, and it was given in \cite{M} a simple model of this kind with the phase-space dimension as same as the Hilbert space dimension of the quantum system. Then, a natural problem is, as raised in \cite{M}, whether or not does there exist a classical theory whose phase-space dimension is much smaller than the Hilbert space dimension for any quantum system. Surprisingly, we show that the answer to this question is negative for any quantum system whose Hilbert space dimension is greater than $2.$

For our purpose, it is sufficient to consider projections for events and trace-one projectors for pure states. We assume that there is a realistic theory underlying the quantum systems. Let $\Omega$ be a suitable realistic space, spanned by a set of variables $X.$ This space can be more general than the phase space and with higher dimensionality, and no {\it priori} hypothesis on it is introduced. We associate any quantum state $| \psi \rangle$ with the distribution of probability $\rho (X | \psi)$ in $\Omega,$ i.e., when the system is prepared to a state $| \psi \rangle,$ $X$ takes a value with a probability given by $\rho.$ In quantum mechanics, if a von Neumann measurement is performed, the state $| \psi \rangle$ collapses to a state $| \phi \rangle$ with probability $| \langle \psi | \phi \rangle|^2.$ In terms of the realistic theory, the system $\Omega$ with coordinates $X$ has a conditional probability $P (\phi | X)$ to give the even $\phi$ conditioned at $\psi.$ The probability of this event is obtained as \begin{equation}\label{eq:CondProb}| \langle \psi | \phi \rangle|^2 = \mathrm{Tr} [\hat{P}_{\phi} \hat{P}_{\psi}] = \int_{\Omega}d X P (\phi | X) \rho (X | \psi),\end{equation}where $\hat{P}_{\phi} = |\phi \rangle \langle \phi |.$ As noted in \cite{M}, the functions $\rho (X | \psi)$ and $P (\phi | X)$ have to satisfy the following conditions:
\begin{equation}\label{eq:PP} \rho (X | \psi) \geq 0,\end{equation}
\begin{equation}\label{eq:ProbUnit} \int_{\Omega}d X \rho (X | \psi) =1,\end{equation}
\begin{equation}\label{eq:CondPP}0 \leq P (\phi | X) \leq 1,\end{equation}for arbitrary states $| \psi \rangle$ and $| \phi \rangle.$

The simplest example that properties \eqref{eq:CondProb}-\eqref{eq:CondPP} are satisfied is \cite{M}:
\begin{equation}\label{eq:PP1} \rho ( \chi | \psi) = \delta (\chi - \psi),\end{equation}
\begin{equation}\label{eq:CondPP1}P (\phi | \chi ) = |\langle \phi| \chi \rangle |^2,\end{equation}
for arbitrary states $| \psi \rangle$ and $| \phi \rangle,$ where $\chi$ is the variable of the realistic space, i.e., all states in the Hilbert space of the quantum system. This example seems sound trivial. However, surprisingly we will prove that the property \eqref{eq:CondPP1} has to hold by assuming that the conditions \eqref{eq:CondProb}-\eqref{eq:CondPP} are fulfilled for every $\psi$ and $\phi$ of a quantum system whose Hilbert space dimension is greater than $2.$ This is the main result of our work. We note that the conditional probabilities \eqref{eq:CondPP1} are not dispersion-free, i.e., the realistic variables cannot fix exactly the results of measurements. This dispersion-free characteristic corresponds to have conditional probabilities equal to $0$ or $1.$

On the other hand, for the case of a complex two-dimensional Hilbert space associated with the spin-$\frac{1}{2}$ system, there are at least two classical hidden variable models, such as Bell's model \cite{Bell1} and Kochen-Specker's model \cite{KS}, which satisfy \eqref{eq:CondProb}-\eqref{eq:CondPP}. Indeed, these two models satisfy Kolmogorov's axiomatic system of classical probability theory, in which not only is the probability nonnegative but also the conditional probabilities are dispersion-free (for details see \cite{BC}). As we will show in the sequel, this is possible just in the case of two-dimensional Hilbert space, where Gleason's theorem \cite{G} does not apply.

For the theorem proof, it is sufficient to consider measurements with trace-one projectors, in this case Eq.\eqref{eq:CondPP} is satisfied because of $\langle \phi | \phi \rangle =1.$ In general, under the addition law of probability a nonnegative probability distribution of a quantum system whose Hilbert space dimension is greater than $2$ can be obtained in terms of positive operator-valued measurements (POVM) by Gleason's theorem \cite{G}. Clearly, Eq.\eqref{eq:CondProb} implies that the addition law of probability holds. Hence, for arbitrary states $| \psi \rangle$ and $| \phi \rangle$ we can define the associated conditional probabilities with respect to a variable $X$ with $\rho (X | \psi) \not= 0$ as
\begin{equation}\label{eq:CondPP2} P (\phi | X) = \mathrm{Tr} [ \hat{B} (X) \hat{P}_{\phi}],\end{equation}
where $\hat{B}(X)$ is a generic Hermitian matrix which depends on $X.$ The realistic variables $X$ can be a vector of continuous and/or discrete variables. $P (\phi | X)$ must satisfy the property \eqref{eq:CondPP}. This implies that
\begin{equation}\label{eq:CondPP3}0 \leq \hat{B} ( X) \leq 1.\end{equation}
Then, assuming \eqref{eq:CondProb}-\eqref{eq:CondPP} and \eqref{eq:CondPP2} as our starting hypotheses, we can prove the theorem.

Indeed, from Eqs. \eqref{eq:CondProb} and \eqref{eq:CondPP2} we have
\begin{equation*} \mathrm{Tr} \left \{ \Big [ \int d X \hat{B} (X) \rho (X | \psi) - \hat{P}_{\psi} \Big ] \hat{P}_{\phi} \right \} = 0,\end{equation*}
for arbitrary states $| \psi \rangle$ and $| \phi \rangle.$ Thus,
\begin{equation}\label{eq:CondPP4}\hat{P}_{\psi} = \int d X \hat{B} (X) \rho (X | \psi).\end{equation}
Since $\rho (X | \psi) \geq 0$ and $0 \leq \hat{B} ( X) \leq 1,$ it is evident that
\begin{equation}\label{eq:CondPP5}\rho (X | \psi) \not= 0 \Rightarrow  \hat{B} ( X) = \lambda (X) \hat{P}_{\psi}, 0 \leq \lambda (X) \leq 1.\end{equation}
Moreover, from \eqref{eq:CondPP4} we have
\begin{equation}\label{eq:CondPP6}\int d X \lambda (X) \rho (X | \psi) = 1.\end{equation}
Then, by \eqref{eq:ProbUnit} we have $\lambda (X) = 1$ whenever $\rho (X | \psi) \not= 0.$ Hence, by \eqref{eq:CondPP2} we have
\begin{equation}\label{eq:CondPP7}\rho (X | \psi) \not= 0 \Rightarrow P (\phi | X) = |\langle \phi| \psi \rangle |^2 .\end{equation}
Also, form \eqref{eq:CondPP5} we conclude that if $\rho (X | \psi) \not= 0$ and $\rho (X | \psi') \not= 0$ then $\psi = \psi'.$ We can define the following one-valued function:
\begin{equation*}\label{eq:fun}\chi: X \to \chi (X) \; \text{such that}\; \rho (X | \chi (X)) \not= 0.\end{equation*}
The function $\chi (X)$ spans the entire Hilbert space, but it is not necessarily invertible. This yields that the dimension of $\Omega$ of the realistic variable space is greater than or equal to the Hilbert space dimension of the quantum system.

In summary, we have the following theorem:

{\it Theorem} For any quantum system whose Hilbert space dimension is greater than $2,$ if there exists a realistic variable model such that the conditions \eqref{eq:CondProb}-\eqref{eq:CondPP} hold, then for arbitrary states $| \psi \rangle$ and $| \phi \rangle$ and any realistic variable $X$ with $\rho (X | \psi) \not= 0,$ one has \begin{equation}\label{eq:CondPP8}P (\phi | X) = |\langle \phi| \psi \rangle |^2,\end{equation} and the dimension of the space $\Omega$ of realistic variables is greater than or equal to the Hilbert space dimension.

Since the properties Eqs.\eqref{eq:CondProb} and \eqref{eq:ProbUnit} have to be satisfied for any realistic theory, in order to obtain a realistic model which uses the mathematical tools distinct from the wave function, we have to discard the properties \eqref{eq:PP} and/or \eqref{eq:CondPP}. As shown in \cite{CEMMMS}, this is sufficient for a realistic theory of a quantum system, no matter the system is continuous or finite.

However, when part of measurements are involved for a quantum system \cite{GM} it is possible to construct a probability distribution associated with every density operator, which satisfies Kolmogorov's axiomatic system of classical probability theory with the phase-space dimension much smaller than the Hilbert-space dimension. The Bohm mechanics provides such a simple example when we only perform measurements of position to evaluate the probability distribution of the position of a particle. Indeed, if the quantum system is in a pure state $\psi,$ the probability to have a particle in the spatial region $\Omega$ is\begin{equation}\label{eq:BohmProb}
\int d \vec{x} P(\Omega | \vec{x}) |\psi (\vec{x}) |^2,\end{equation}where \begin{equation}\label{eq:BohmCondProb}
P(\Omega | \vec{x}) = \int_{\Omega} \delta ( \vec{x}'- \vec{x} ) d \vec{x}',\end{equation}$\delta ( \vec{x}'- \vec{x} )$ being the conditional probability density of finding the particle at $\vec{x}'.$

On the other hand, when we only consider certain part of measurements for a quantum system it depends on the state of the system whether or not there exists a classical probability description of it. This is the starting point of Bell's inequality \cite{Bell}, which was designed to rule out various kinds of local hidden variable theories based on EPR's notion of local realism \cite{EPR}. To this end, we consider the Clauser-Horne-Shimony-Holt (CHSH) inequality \cite{CHSH}. Let us consider a system of two qubits labelled by $1$ and $2.$ Let $A, A'$ denote spin observables on the first qubit, and $B, B'$ on the second one. We write $AB,$ etc., as shorthand for $A \otimes B$ and $\langle AB \rangle_{\psi}: = \langle \psi | AB | \psi \rangle$ for the expectation of $AB$ in the state $\psi.$ When only are spin measurements performed on two qubits, if there exists a classical probability description for the system then the following CHSH inequality holds:\begin{equation}\label{CHSH} \langle a b\rangle + \langle a b'\rangle + \langle a' b\rangle - \langle a' b' \rangle \leq 2,\end{equation}
where $a^{(')}$ denotes the corresponding classical quantity of $A^{(')}$ on the first qubit, and $b^{(')}$ on the second one. For any product state Eq.\eqref{CHSH} holds, thus a classical probability description of the system on a product state is possible if spin measurements are involved. However, for some entangled states, such as Bell's states\begin{equation}\label{BellState}| \phi^{\pm} \rangle = \frac{1}{\sqrt{2}} (| \uparrow \uparrow \rangle \pm | \downarrow \downarrow \rangle),\; | \psi^{\pm} \rangle = \frac{1}{\sqrt{2}} (| \uparrow \downarrow \rangle \pm | \downarrow \uparrow \rangle),\end{equation} Eq.\eqref{CHSH} can be violated when measurements of some spin observales $A, A'$ on the first qubit and $B, B'$ on the second one are performed. (For details on Bell inequalities and entanglement see the review paper \cite{WW}.) This immediately concludes that there is no classical probability description for the system on such entangled states.

It is well known that the dimension of the Hilbert space grows exponentially with the physical size of the system. For example, it is $2^N$ for $N$ spin-$1/2$ particles. Every known classical variable theory has this feature. Our theorem shows that any classical theory for quantum systems whose Hilbert space dimension is greater than $2$ has to have this feature. As is well known, the Bohm mechanics provides a classical description of quantum systems and uses the same mathematical tools of the standard approach as the wave functions, according to our theorem, it is thus one of the most basic classical theory for quantum mechanics. Recently, the theory has been extended to quantum fields and accounts for creation and annihilation of a particle \cite{DGTZ}.

However, if we discard some required property, such as positivity, we can obtain a real-variable theory whose phase-space dimension can be considerably reduced. For example, the Hilbert space dimension of one boson mode is infinite, since the number of particles goes from zero to infinity, but the corresponding Wigner functions have only two variables, although we cannot regard these variables as describing a classical system because of the negativity of the Wigner function. Also, for a quantum system of a $N$-dimensional Hilbert space the domain of definition of Wigner-Weyl representatives in \cite{CEMMMS} is exactly the `classical' discrete phase-space lattice of just $N^2$ points with $0 \leq q,p \leq N-1.$

The original Wigner and $Q$ distributions were developed for continuous systems. In recent years various phase space and other quasi-probability representations of finite-dimensional quantum systems have been proposed (for a recent review see \cite{V}). The term quasi-probability refers to the fact that the function is not a true density as it takes on negative values for some quantum states. Such representations have provided insight into fundamental structures for finite-dimensional quantum systems. For example, the representation proposed by Wootters identifies sets of mutually unbiased bases \cite{WGH}. Inspired by quantum information and computation \cite{NC}, there has been growing interest in the application of discrete phase-space representation to analyze the quantum-classical contrast for finite-dimensional systems \cite{PMLGC}. In particular, a physically reasonable Wigner quasi-probability with a discrete phase-space of finitely many points is set up for the finite dimensional Hilbert space \cite{CEMMMS} by drawing inspiration from Dirac's work on functions of non-commuting observables. Accordingly, we raise the following question: Does a real-variable theory with a discrete phase-space of finitely many points there exist for any state of a finite dimensional system satisfying the properties \eqref{eq:CondProb}-\eqref{eq:ProbUnit}? It is known that if one asks for a phase-space description of a quantum state obeying a small number of very reasonable conditions, then the Wigner function is the unique answer. Hence, there is no evident reason for assuming there is a nonnegative probability distribution for a quantum state. The introduction of a probability distribution $\rho (X|\psi)$ and a conditional probability $P(\phi |X)$ for each even allows one to place the nature of fundamental structures of finite-dimensional quantum systems onto a clear and well-defined ground, and the proof of whether or not there exists a nonnegative quasi-probability with a discrete phase-space of finitely many points for any state of finite quantum systems would cast new light upon this question.

In conclusion, we have studied the possibility of a classical description of quantum states and events in terms of probability distributions and conditional probabilities that must satisfy the properties \eqref{eq:CondProb}-\eqref{eq:CondPP}. We demonstrated that the distributions and/or conditional probabilities have to assume negative values, except for a simple model whose realistic space dimension is not smaller than the Hilbert space dimension of the quantum system. As shown in \cite{CEMMMS}, discarding the nonnegative of probability and/or conditional probabilities we can always obtain a realistic theory for any quantum system, no matter the system is continuous or finite. Therefore, for any quantum system whose Hilbert space dimension is greater than $2,$ the negativity of probability and/or conditional probabilities is necessary and sufficient for a realistic theory with much lower dimensionality. This gives a negative answer to a question proposed by Montina \cite{M} whether or not does there exist a classical theory whose phase-space dimension is much smaller than the Hilbert space dimension for any quantum system. Finally, we discuss the real-variable theory for finite-dimensional quantum systems and conclude with a nontrivial question: Does a real-variable theory with a discrete phase-space of finitely many points exist whose probability distribution is nonnegative for any state of finite-dimensional quantum systems? It seems impossible to develop a $Q$ distribution for any finite dimensional system, since there is no analogue of coherent states for the finite dimensional Hilbert spaces, whereas a real-variable theory with a discrete phase-space of finitely many points and nonnegative probability for any state of finite-dimensional quantum systems would have important implication in quantum information and quantum computation. The solution to this problem could clarify the nature of the exponential speedup of quantum computers \cite{SG}.

This work was partially supported by the National Natural Science Foundation of China under Grant No.10775175.

\bibliography{apssamp}

\end{document}